\RequirePackage{lineno}
\documentclass[preprint,aps,superscriptaddress]{revtex4}

\usepackage{graphicx}
\usepackage[caption=false]{subfig}
\usepackage{epsfig}
\usepackage{dcolumn}
\usepackage{bm}
\usepackage{ulem}
\usepackage{xcolor}
\usepackage{textcomp}
\usepackage{indentfirst}
\usepackage{epstopdf}
\usepackage{amsmath}
\usepackage{natbib}
\usepackage[colorlinks,linkcolor=blue,anchorcolor=blue,citecolor=blue]{hyperref}
\usepackage[all]{hypcap}
\usepackage{cleveref}

\begin{document}

\title{Ultrafast Electronic Dynamics of a Weyl Semimetal MoTe$_2$ Revealed by Time and Angle Resolved Photoemission Spectroscopy}

\author{Guoliang Wan}
\altaffiliation{These authors contributed equally to this work.}
\affiliation{State Key Laboratory of Low Dimensional Quantum Physics and Department of Physics, Tsinghua University, Beijing 100084, China}

\author{Wei Yao}
\altaffiliation{These authors contributed equally to this work.}
\affiliation{State Key Laboratory of Low Dimensional Quantum Physics and Department of Physics, Tsinghua University, Beijing 100084, China}

\author{Kenan Zhang}
\affiliation{State Key Laboratory of Low Dimensional Quantum Physics and Department of Physics, Tsinghua University, Beijing 100084, China}

\author{Changhua Bao}
\affiliation{State Key Laboratory of Low Dimensional Quantum Physics and Department of Physics, Tsinghua University, Beijing 100084, China}

\author{Hongyun Zhang}
\affiliation{State Key Laboratory of Low Dimensional Quantum Physics and Department of Physics, Tsinghua University, Beijing 100084, China}

\author{Haijun Zhang}
\affiliation{National Laboratory of Solid State Microstructures and School of Physics,  Nanjing University, Nanjing 210093, China}
\affiliation{Collaborative Innovation Center of Advanced Microstructures, Nanjing, China}

\author{Yang Wu}
\affiliation{Department of Physics and Tsinghua-Foxconn Nanotechnology Research Center, Tsinghua University, Beijing, 100084, China}

\author{Shuyun Zhou}
\altaffiliation{Correspondence should be sent to \href{mailto:syzhou@mail.tsinghua.edu.cn}{syzhou@mail.tsinghua.edu.cn}}
\affiliation{State Key Laboratory of Low Dimensional Quantum Physics and Department of Physics, Tsinghua University, Beijing 100084, China}
\affiliation{Collaborative Innovation Center of Quantum Matter, Beijing, P.R. China}

\date{\today}

\begin{abstract}

{\bf A Weyl semimetal is a new type of topological quantum phase with intriguing physics near the Weyl nodes. Although the equilibrium state of Weyl semimetals has been investigated, the ultrafast dynamics near the Weyl node in the nonequilibrium state is still missing. Here by performing time and angle resolved photoemission spectroscopy on type-II Weyl semimetal MoTe$_2$, we reveal the dispersion of the unoccupied states and identify the Weyl node at 70 meV above E$_F$. Moreover, by tracking the ultrafast relaxation dynamics near the Weyl node upon photo-excitation with energy, momentum and temporal resolution, two intrinsic recovery timescales are observed, a fast one of 430 fs and a slow one of 4.1 ps, which are associated with hot electron cooling by optical phonon cascade emission and anharmonic decay of hot optical phonons respectively. The electron population shows a metallic response, and the two temperature model fitting of the transient electronic temperature gives an electron-phonon coupling constant of $\lambda\langle\Omega^2\rangle\simeq32$ $\textrm{meV}^2$. Our work provides important dynamic information for understanding the relaxation mechanism of a Weyl semimetal and for exploiting potential applications using ultrafast optical control.}
\end{abstract}

\maketitle

\section{INTRODUCTION}

Three-dimensional (3D) Weyl semimetal \cite{WanXG,WeylRMP} is a novel topological phase hosting 3D Weyl fermions originally proposed in high-energy physics. Their electronic structures are characterized by conical dispersion near the Weyl nodes and open Fermi arcs connecting Weyl nodes with opposite chiralities \cite{xu2015discovery,lv2015experimental,FangZTaAs}. Weyl semimetals exhibit novel quantum phenomena such as chiral anomaly \cite{ChenGFPRX,JiaSNatComm2016} and interesting optical properties governed by chirality, e.g., light induced photocurrent \cite{ishizuka2016PRL,chan2017photocurrents,de2017QPGE,ma2017WeylChirality}, giant anisotropic nonlinear optical response \cite{OrensteinSHG} and photoinduced topological phase transition \cite{chan2016Weyl,hubener2017FloquetWSM}. Moreover, it has been reported that the Weyl nodes can be selectively excited and controlled by circularly polarized light \cite{DaiXTRARPES,ma2017WeylChirality}. Such selective control of the Weyl nodes may provide new opportunities for device applications in analogy to the selective control of valley polarizations in valleytronics \cite{CuiXDValley,HeinzValley,FengJ}. Understanding the dynamic evolution of photo-excited carriers near the Weyl node is critical for revealing the intriguing physics and for exploiting potential applications using ultrafast optical control \cite{zhu2017Cd3As2}. So far experimental research on Weyl semimetals has been focused on the equilibrium states, while dynamic information about the relaxation of charge carriers upon transient photo-excitation is still missing.

Recently, type-II Weyl semimetal with Weyl Fermions emerging at the topologically protected touching points between electron and hole pockets has been proposed \cite{soluyanov2015type} and realized experimentally in the T$_d$ phase of MoTe$_2$ \cite{kedeng2016MoTe2,huang2016spectroscopic}. Signatures of the topological Fermi arcs have been revealed by angle resolved photoemission spectroscopy (ARPES) \cite{kedeng2016MoTe2,huang2016spectroscopic} and quasi-particle interference (QPI) pattern in scanning tunneling microscopy (STM) \cite{kedeng2016MoTe2,zheng2016atomic,russmann2017STM}. However, the Weyl nodes are above the Fermi energy E$_F$, making it challenging for conventional ARPES to resolve the dispersion near the Weyl node. Here by performing time and angle resolved photoemission spectroscopy (TR-ARPES), we reveal the electronic dispersion near the Weyl node and confirm the Weyl node at $\approx$ 70 meV above E$_F$. Furthermore, by tracking the nonequilibrium states in the momentum space upon photo-excitation, we reveal the ultrafast relaxation dynamics near the Weyl node. Two intrinsic timescales are observed in the relaxation process with 430 femtoseconds (fs) and 4.1 picoseconds (ps), indicating two distinct relaxation channels. In addition, the electron population shows a metallic response and the two temperature model (TTM) fitting gives the second moment of Eliashberg electron-phonon ($el$-$ph$) coupling function, $\lambda\langle\Omega^2\rangle\simeq$ 32 $\pm$ 1.2 $\textrm{meV}^2$. Our work provides important information for understanding the ultrafast dynamics of a photo-excited Weyl semimetal.

\section{EXPERIMENTAL RESULTS}

\begin{figure*}
	\subfloat{\label{fig1a}}
	\subfloat{\label{fig1b}}
	\subfloat{\label{fig1c}}
	\subfloat{\label{fig1d}}
	\subfloat{\label{fig1e}}
	\subfloat{\label{fig1f}}
	\subfloat{\label{fig1g}}
	\subfloat{\label{fig1h}}
	\subfloat{\label{fig1i}}
	\subfloat{\label{fig1j}}
	\subfloat{\label{fig1k}}
	\subfloat{\label{fig1l}}
	\subfloat{\label{fig1m}}
	\subfloat{\label{fig1n}}
	\includegraphics[width=16.5 cm] {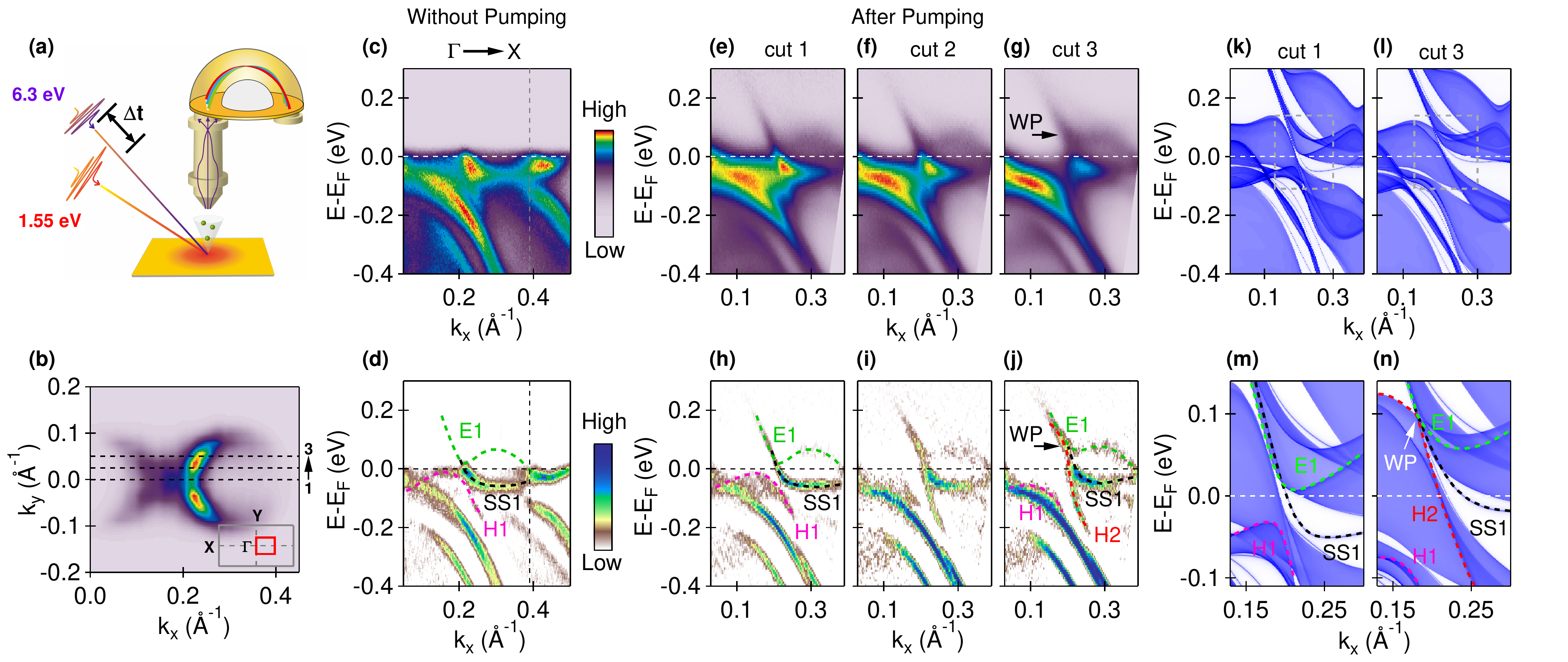}
	\caption{{\bf Unoccupied states above E$_F$ and the Weyl node in T$_d$-MoTe$_2$ revealed by TR-ARPES.} \textbf{(a)} Schematic of TR-ARPES experimental setup.  \textbf{(b)} Fermi surface map measured at 32.5 eV photon energy.  The red rectangle in the inset shows the measured range inside the projected Brillouin zone. \textbf{(c-d)} ARPES spectrum and corresponding curvature plot along the $\Gamma$-X direction measured with 6.3 eV laser without pumping.  \textbf{(e-g)} Transient ARPES spectra measured at cut 1 to cut 3 indicated by dashed black lines in panel (b) after pumping. \textbf{(h-j)} Corresponding curvature plots to panels (e-g). Dashed lines are guides to the eyes for the band dispersions.  The Weyl node is marked by the black arrow in panel (j). \textbf{(k-l)} Calculated spectral weight for cut 1 and cut 3. \textbf{(m-n)} Zoom-in of the dashed rectangular regions in panels (k-l). The arrow points to the Weyl node.}
	\label{fig1}
\end{figure*}

Figure \ref{fig1a} shows a schematic illustration of our TR-ARPES experimental setup. A pump pulse of 1.55 eV  photon energy is used to excite the sample to a nonequilibrium state, and subsequently a probe pulse of 6.3 eV photon energy is used to capture the transient electronic structures at different time delays. By performing ARPES measurements in the pump-probe scheme, TR-ARPES can map out dispersions of the unoccupied states \cite{ShenBi2Se3PRL2013}. Moreover, by changing the time delay between the pump and probe pulses, we can directly capture the dynamic evolution of photo-excited charge carriers with energy, momentum and time resolved information \cite{Lanzara2011NPhy} and thus disentangle the different relaxation processes in the time domain.

TR-ARPES allows to measure the electronic structure near the Weyl node, which is above E$_F$ and not accessible using conventional ARPES. Figure \ref{fig1b} shows the Fermi surface map of T$_d$-MoTe$_2$ and Figure \ref{fig1c} shows the measured ARPES dispersion along the $\Gamma$-X direction (cut 1 in Fig.~\ref{fig1b}) without pumping. Figure~\ref{fig1d} shows the corresponding curvature plot, which has been applied to the ARPES data to enhance visualization of the dispersion \cite{DingRsi}. A hole band (H1) and a surface state (SS1) are observed, while the portion of conduction band (E1) above E$_F$ is undetectable. Figure \ref{fig1e} and \ref{fig1h} show the TR-ARPES transient image and curvature plot after photo-excitation, where the entire W-shaped conduction band E1 becomes observable. In order to search for the signature of Weyl node at the boundary between the electron and hole pockets, the evolution of these bands is shown in Fig.~\ref{fig1e}-\ref{fig1g} with the corresponding curvature plots in Fig.~\ref{fig1h}-\ref{fig1j}. From cut 1 to cut 3, the small hole band (H1) shrinks and an additional hole band (H2) emerges. This hole band gradually approaches the electron band E1, and they eventually touch at the Weyl node (indicated by arrows in Fig.~\ref{fig1g} and \ref{fig1j}), in agreement with main features in the calculated spectra shown in Fig.~\ref{fig1k}-\ref{fig1l} and zoom-in images (Fig.~\ref{fig1m}-\ref{fig1n}). By using TR-ARPES, we directly identify the location of the Weyl node to be at 70 $\pm$ 15 meV above E$_F$, and further establishes T$_d$-MoTe$_2$ as a type-II Weyl semimetal phase.

\begin{figure*}
	\subfloat{\label{fig2a}}
	\subfloat{\label{fig2b}}
	\subfloat{\label{fig2c}}
	\subfloat{\label{fig2d}}
	\subfloat{\label{fig2e}}
	\subfloat{\label{fig2f}}
	\subfloat{\label{fig2g}}
	\subfloat{\label{fig2h}}
	\subfloat{\label{fig2i}}
	\subfloat{\label{fig2j}}
	\includegraphics[width=14 cm] {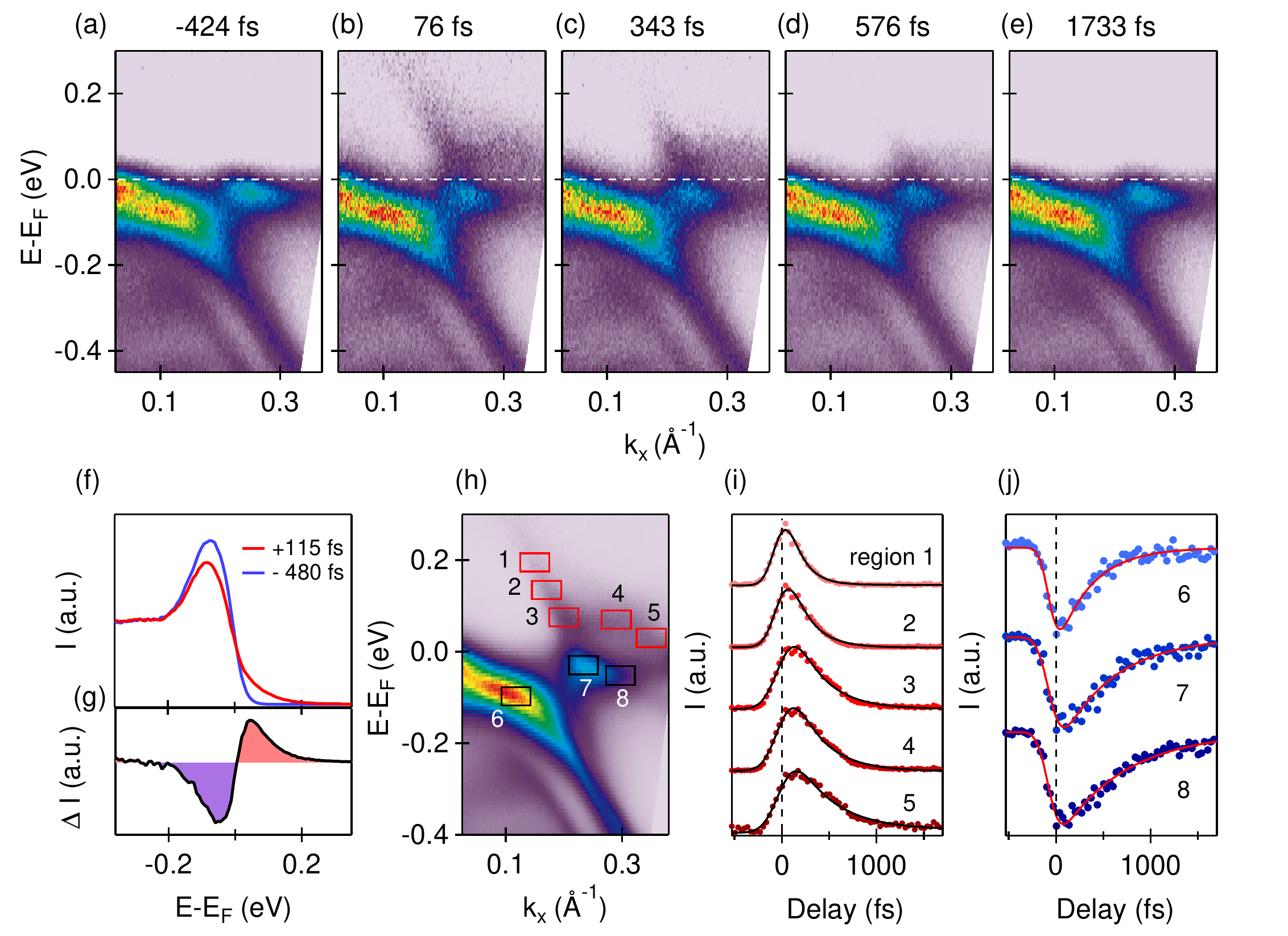}
    \caption{{\bf Dynamic evolution of the transient occupation along $\Gamma$-X direction after optical pumping.} \textbf{(a-e)} TR-ARPES
spectra taken at selected pump-probe delays along cut 3 in Fig.~1 at a pump fluence of 1.6 $\textrm{mJ/cm}^2$. \textbf{(f-g)} Transient photoemission intensity integrated over the momentum range as a function of energy and the differential spectral weight to the equilibrium state. \textbf{(h)} Transient image recorded at $\Delta$t=120 fs. \textbf{(i-j)} Dynamic evolution of transient photoemission intensity integrated within the rectangles in panel (h). The data are normalized to peak intensities and vertically offset for clarity. The solid lines are exponential fits to the data.}
	\label{fig2}
\end{figure*}

The relaxation dynamics near the Weyl node is revealed by taking snapshots of the transient spectra at different pump-probe delays. Figure \ref{fig2a} shows data taken at negative delay, i.e., before the pump excitation. Upon photo-excitation, electrons are immediately populated into the electronic states above E$_F$ (Fig.~\ref{fig2b}), and subsequently relax into the equilibrium state within a few hundred fs (Fig.~\ref{fig2c}-\ref{fig2e}). Figure \ref{fig2f} and \ref{fig2g} show the transient momentum-integrated intensity curve at 115 fs and -480 fs delay and the differential intensity plot. A dramatic increase and decrease of intensity are clearly observed within 200 meV above and below E$_F$, respectively. To further investigate the energy dependence, we show in Fig.~\ref{fig2i} and \ref{fig2j} the temporal evolution of transient photoemission intensity for the conduction band and valence bands at energy and momentum positions marked in Fig.~\ref{fig2h}. For the W-shaped conduction band (E1), higher energy electrons (region 1 in Fig.~\ref{fig2i}) relax faster while lower energy electrons near E$_F$ (region 5) survive for slightly longer time. The rising edge shifting toward positive time delay from high energy to low energy in Fig.~\ref{fig2i} is usually caused by the cascade relaxation processes \cite{ShenBi2Se32012}, in which electrons relax from high energy states to low energy states in sequence. These observations suggest that the relaxation is mainly through intraband scattering. Similar trend with longer relaxation time for states near E$_F$ is also observed in the valence band (Fig.~\ref{fig2j}). The recovery process can be further fitted using an empirical formula by including an error function for the rising edge and an exponential decay, convoluted with a Gaussian function to account for the temporal resolution. The extracted relaxation time $\tau$ in Fig.~\ref{fig2i} varies from 150 fs at high energy to more than 400 fs at low energy. Such recovery time is much faster than the few ps observed in topological insulator Bi$_2$Se$_3$ \cite{ShenBi2Se32012,GedikBi2Se3PRL2012}, ZrTe$_5$ \cite{manzoni2015ZrTe5} and Bismuth \cite{PerfettiBi2012}. Since hot electrons can cool down on a sub-picosecond timescale by coupling with optical phonons as reported in other materials \cite{kampfrath2005PRL,Perfetti2007,Hofmann2013}, we tentatively attribute the observed ultrafast electron relaxation process to the intraband scattering mediated by $el$-$ph$ interaction.

\begin{figure*}
	\subfloat{\label{fig3a}}
	\subfloat{\label{fig3b}}
	\subfloat{\label{fig3c}}
	\subfloat{\label{fig3d}}
	\subfloat{\label{fig3e}}
	\subfloat{\label{fig3f}}
	\subfloat{\label{fig3g}}
	\subfloat{\label{fig3h}}
	\subfloat{\label{fig3i}}
	\subfloat{\label{fig3j}}
	\subfloat{\label{fig3k}}
	\subfloat{\label{fig3l}}
	\subfloat{\label{fig3m}}
	\subfloat{\label{fig3n}}
	\includegraphics[width=16.5 cm] {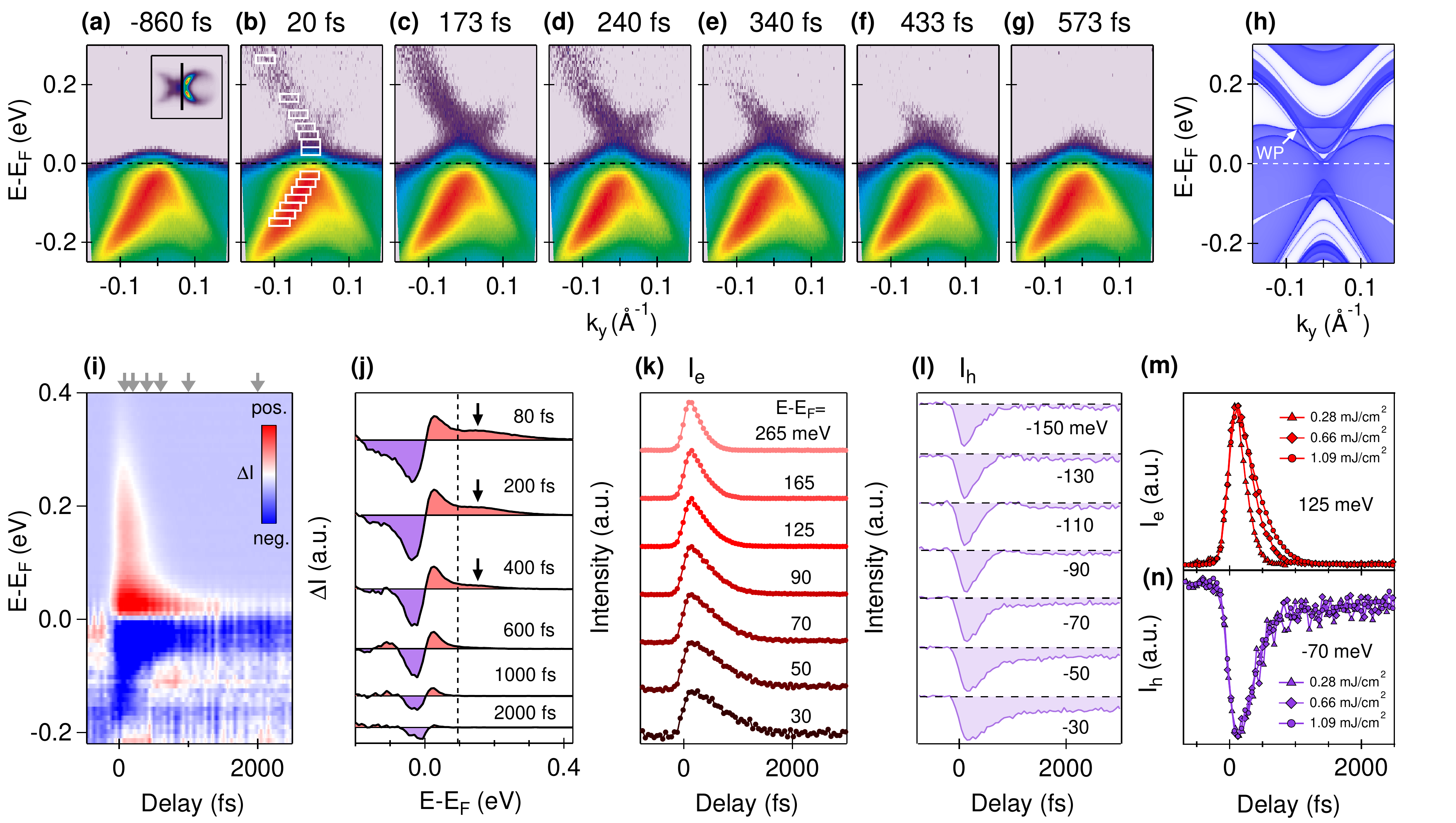}
    \caption{{\bf Relaxation dynamics for electrons and holes near the Weyl node after photo-excitation.} \textbf{(a-g)}
Snapshots of transient occupation at different time delays along the cut through the Weyl node indicated by solid line in the inset of panel (a) at a pump fluence of 0.66 $\textrm{mJ/cm}^2$. \textbf{(h)} Corresponding calculated spectral intensity plot. \textbf{(i)} Momentum-integrated photoemission differential intensity versus energy and time delay by subtracting the equilibrium intensity. \textbf{(j)} Differential energy distribution curves at selected time delays extracted from panel (i). \textbf{(k-l)} Temporal evolution of transient photoemission intensity for photo-excited electrons and holes integrated within the rectangles in panel (b). \textbf{(m-n)} Pump fluence dependence in the population relaxation dynamics of photo-excited electrons at 125 meV and holes at -70 meV respectively.}
	\label{fig3}
\end{figure*}

To further reveal the relaxation mechanism of photo-excited carriers near the Weyl node, we show in Fig.~\ref{fig3a}-\ref{fig3g} snapshots of the band structure through the Weyl node measured parallel to $\Gamma$-Y direction at different delays, in which simpler band dispersion can be resolved. A conical dispersion similar to the theoretical calculation (Fig.~\ref{fig3h}) is observed after transient photo-excitation and relaxes within a few hundred fs. The timescale is consistent with that measured along the $\Gamma$-X direction. Figure \ref{fig3i} shows the momentum-integrated differential intensity relative to negative delay as a function of energy and delay. The relaxation dynamics is strongly dependent on energy. Within the first few hundred fs after photo-excitation, two relaxation components with one near E$_F$ (bright red in Fig.~\ref{fig3i}) and one extending to higher energies (light red) are clearly observed. Figure \ref{fig3j} shows the selected differential energy profiles extracted from Fig.~\ref{fig3i}. For the population gains above E$_F$, in addition to the peak as commonly seen in pump-probe measurements, there is a hump indicated by arrows between 0.1 and 0.3 eV within $\sim$ 400 fs, which further confirms the existence of the second component. At delay beyond 400 fs, the fast component almost disappears and the nonvanishing differential intensity between -0.1 eV to 0.1 eV decays with the slow component. Therefore there are two contributions to the population relaxation dominated at different timescale and at different energy range.

The relaxation dynamics of photo-excited electrons and holes can be resolved separately with both energy and momentum resolved information. We first focus on the evolution of photo-excited electrons in the conduction band shown in Fig.~\ref{fig3k}.  Above 90 meV, the relaxation time is fast (within 200 fs), while below 90 meV, it shows pronounced energy dependent and quickly increases from $\sim$ 200 fs at 90 meV to $\sim$ 500 fs at 30 meV. A more notable change is also observed in the dynamics of photo-excited holes shown in Fig.~\ref{fig3l}. At energy below -90 meV, the relaxation occurs within $\sim$ 200 fs and is almost energy independent, while above -90 meV, an additional slower relaxation component extending up to several ps is observed. A similar abrupt change in the energy dependence of population dynamics has been also reported in optimally-doped Bi-2212 \cite{ZXShen2015BSCCO}. The photo-excited electrons and holes also show different pump fluence dependence. The relaxation time for photo-excited electrons increases with fluence (Fig.~\ref{fig3m}). In contrast to electrons, photo-excited holes in the valence band remains almost fluence independent (Fig.~\ref{fig3n}, see more details in Fig.~S2 in the supplementary information). Such distinctive relaxation dynamics suggest strong electron-hole asymmetry. A possible reason is the extremely low density of states near the top of the valence band, which hinders the electronic relaxation from conduction band to valence band and results in a nearly fluence independent recovery time.

\begin{figure*}
	\subfloat{\label{fig4a}}
	\subfloat{\label{fig4b}}
	\subfloat{\label{fig4c}}
	\subfloat{\label{fig4d}}
	\subfloat{\label{fig4e}}
    \subfloat{\label{fig4f}}
	\includegraphics[width=14.5 cm] {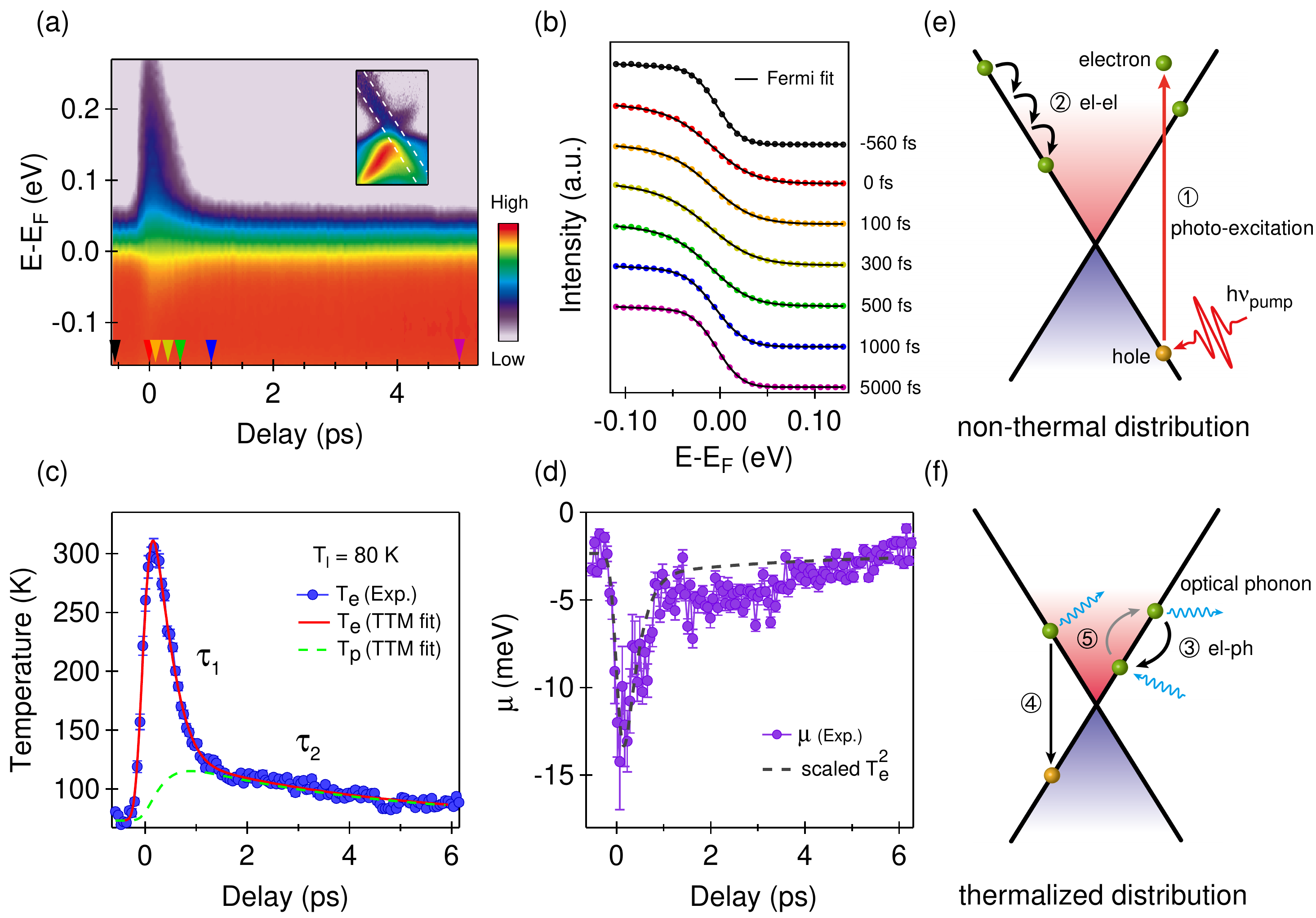}
	\caption{{\bf Electronic cooling by generation of nonequilibrium phonons.} \textbf{(a)} Temporal evolution of energy distribution curves for selected band by integrating over the momentum range marked in the inset. \textbf{(b)} Selected curves at different delays extracted from panel (a). The solid black lines are fitting curves to extract the electronic temperature $T_e$ and chemical potential $\mu$. \textbf{(c-d)} Extracted $T_e$ and $\mu$ as a function of delay at lattice temperature $T_l$ = 80 K. The vertical error bars are the fitted standard deviations. The solid red line and dashed green line in panel (c) are two temperature model simulations of hot electron and hot optical phonon temperature. $\tau_1$ and $\tau_2$ are two characteristic time scales, corresponding to the electronic cooling and hot phonon anharmonic decay, respectively. The broken line in panel (d) is the scaled curve plot of $T_e^2$. \textbf{(e-f)} Schematics of carrier and phonon excitation and relaxation process. After thermalization by carrier-carrier scattering, electrons can cool down through phonon emission (indicated by wiggled arrows).}
	\label{fig4}
\end{figure*}

In addition to the transient populations discussed above, we further extract the effective electronic temperature by analyzing the integrated spectral intensity which is effectively the transient statistical distribution of the carriers along the dispersing band \cite{GedikBi2Se3PRL2012,gierz2013,Hofmann2013}. Figure \ref{fig4a} shows the temporal evolution of momentum-integrated intensity for the conical dispersion (region marked by white broken lines in the inset). At all time delays, the electron distribution can be well fitted by Fermi-Dirac distribution convoluted with a Gaussian resolution function (Fig.~\ref{fig4b}), indicating a metallic response as observed in graphene \cite{Hofmann2013}. This suggests that thermalization of hot electrons occurs on an extremely short time scale within our temporal resolution. Such thermal-equilibrium distribution allows us to extract an effective electronic temperature $T_e$ and chemical potential $\mu$, as shown in Fig.~\ref{fig4c}-\ref{fig4d}. The maximum electronic temperature $T_{e,max}$ is two times smaller than the estimated value (see supplementary information), suggesting that the spatial heat diffusion is fast in the lattice. Two distinct relaxation components are observed in both $T_e$ and $\mu$, with two characteristic timescales of $\tau_1$= 430 $\pm$ 30 fs and $\tau_2$= 4.1 $\pm$ 0.5 ps, respectively (Fig.~S4 in the supplementary information). These are significant hints for taking into account the hot phonons in the relaxation process \cite{Hofmann2013,Perfetti2007}. The fast recovery component is attributed to hot electron cooling through coupled optical phonons, while the slow component is determined by the anharmonic decay of hot optical phonons. The anharmonic behavior in bulk T$_d$-MoTe$_2$ is also observed by temperature dependent Raman spectroscopy \cite{joshi2016phonon}, in which an optical phonon decay into multiple acoustic phonons.

\section{DISCUSSIONS}

To quantitatively examine the $el$-$ph$ mediated relaxation, we employ a minimal two temperature model (TTM) \cite{allen1987} to fit the electronic temperature. Considering the lattice as a large phonon bath, we assume that the lattice temperature $T_l$ remains at $T_0$= 80 K after photo-excitation, and the TTM model involves two important temperatures: the electronic temperature $T_e$, the hot optical phonon temperature $T_p$. Based on the above assumption, the TTM is simplified into two coupled differential equations:
$$ \frac{dT_e}{dt}= -\frac{3\lambda\langle\Omega^2\rangle}{\pi\hbar k_B}\frac{T_e-T_p}{T_e}+\frac{S(t)}{C_e} \eqno{(1)} $$
$$ \frac{dT_p}{dt}= \frac{C_e}{C_p} \frac{3\lambda\langle\Omega^2\rangle}{\pi\hbar k_B}\frac{T_e-T_p}{T_e}-\frac{T_p-T_0}{\tau_\beta} \eqno{(2)} $$
where $C_e$ and $C_p$ are the specific heat of electrons and phonons; $S(t)$ is the driven term by the pump pulse with a Gaussian profile of FWHM $\sim$ 50 fs and an absorbed energy density of $\sim$ 130 $\textrm{J/cm}^3$; $\lambda$ and $\Omega$ represent the $el$-$ph$ coupling constant and coupled optical phonon energy; $\tau_\beta$ is the decay time of hot nonequilibrium phonons; $k_B$ and $\hbar$ are the Boltzmann and Planck constants, respectively. The TTM simulation of the photo-excited electronic temperature and hot optical phonon temperature is shown as red and green curves in Fig.~\ref{fig4c} respectively. Such good fitting further supports the hot phonon involved relaxation scenario. The average $el$-$ph$ coupling strength $\lambda\langle\Omega^2\rangle$= 32 $\pm$ 1.2 $\textrm{meV}^2$ and the phonon decay time $\tau_\beta$= 3.8 $\pm$ 0.4 ps are extracted from the TTM fitting, where $\lambda\langle\Omega^2\rangle=2\int\Omega\alpha^2F(\Omega)d\Omega$ is the second moment of the Eliashberg spectral function $\alpha^2F(\Omega)$ \cite{allen1987}. Such $\lambda\langle\Omega^2\rangle$ is much smaller than that observed in graphite ($\sim$ 0.028 $\textrm{eV}^2$) \cite{sugawara2007} and Bi-2212 ($\sim$ 360 $\textrm{meV}^2$) \cite{Perfetti2007}. Since the electronic relaxation time $\tau_{\alpha}$ is determined by the $el$-$ph$ coupling strength with $\tau_{\alpha}=(\pi/3)(\hbar k_BT_e/\lambda\langle\Omega^2\rangle)$, this results in a larger electronic cooling time in MoTe$_2$ ($\tau_{\alpha}\simeq$ 430 fs) compared to that in graphite ($\sim$ 150-200 fs) \cite{Hofmann2013,breusing2009} and Bi-2212 ($\sim$ 110 fs) \cite{Perfetti2007}. Furthermore, taking the $el$-$ph$ coupling constant $\lambda\simeq$ 0.3 reported in bulk T$_d$-MoTe$_2$ \cite{takahashi2017EPC} and assuming that the coupling is dominated by one phonon mode, i.e., $\alpha^2F(\Omega)\propto\delta(\Omega-\Omega_0)$, the mean phonon energy coupled to the hot electrons is estimated to be $\Omega_0$ $\sim$ 10.3 meV. This energy is close to the $A_1$ optical phonon mode vibrating in the ac-plane with a calculated wavenumber of 78 $cm^{-1}$ \cite{zhang2016raman}. For such phonon energy, the $el$-$ph$ scattering phase space is vanishingly small owing to the accessible electronic structure in the relaxation process, while acoustic phonons need a larger momentum window at such energy \cite{chen2016Raman} and thus excludes acoustic phonon emission as the dominant cooling mechanism. Consequently, photo-excited hot electrons cool down within 400 fs through optical phonon cascade relaxation \cite{kovalev1996GaN} and generate a large number of optical phonons. When the temperature of hot electrons and hot phonons reaches quasi-equilibrium, these hot optical phonons can be reabsorbed by electrons, thereby slowing down the hot electron cooling process. Therefore, the relaxation of the hot electrons is dominated by the decay of optical phonons, which serves as the hot phonon bottleneck \cite{langot1996GaAs}. Similar behavior is also observed in the chemical potential $\mu$ (Fig.~\ref{fig4d}). After photo-excitation, $\mu$ decreases immediately and recovers as electrons cool down. The changes of $\mu$ can be qualitatively understood as a result of elevated effective electronic temperature $T_e$ by $\Delta\mu=-(\pi^2/12)(k_BT_e)^2/E_F$, as expected in simple metals \cite{ashcroft1976}. The dashed curve in Fig.~\ref{fig4d} is a scaled curve plot of $T_e^2$ in Fig.~\ref{fig4c}, which overall shows similar temporal evolution to $\mu$ in the electron cooling process.

Figure \ref{fig4e} and \ref{fig4f} show a schematic summary of the thermalization and relaxation process. After photo-excitation (process \textcircled{1} in Fig.~\ref{fig4e}) and thermalization through direct electron-electron scattering (Process \textcircled{2}), hot electrons reach a quasi-equilibrium distribution shortly, with an elevated electronic temperature $T_e$. The hot electrons cool down fast by generating a large number of hot optical phonons ($T_p$) through intraband cascade scattering (\textcircled{3} in Fig.~\ref{fig4f}) or interband electron-hole recombination (process \textcircled{4}).  The hot nonequilibrium phonons then dissipate their energy to other lattice vibrations ($T_l$) through anharmonic phonon-phonon interaction, and the reabsorption of hot phonons slows down the electron cooling process, and leads to the slower relaxation component determined by the lifetime of optical phonons (process \textcircled{5} in Fig.~\ref{fig4f}).

\section{CONCLUSIONS}

In summary, we reveal the unoccupied electronic states of T$_d$-MoTe$_2$ and confirm the Weyl node located at energy 70 meV above $E_F$. Moreover, the dynamic evolution of photo-excited electrons and holes is revealed with energy, momentum and time resolution. Two intrinsic relaxation timescales are observed, with the faster component of 430 fs attributed to the hot electron cooling through optical phonon emission, while the slow component of 4.1 ps is determined by the anharmonic decay of hot optical phonons. The $el$-$ph$ coupling strength is extracted by TTM simulation. Our work provides dynamic information on this new type of topological phase, and highlights the significant role of $el$-$ph$ interaction in the relaxation process of quasiparticles near the Weyl node upon photo-excitation.

{\large{\bf Methods}}

High quality MoTe$_2$ single crystals were synthesized by chemical vapor transport (CVT) method as detailed in our previous work \cite{kedeng2016MoTe2,zhang2016raman}. TR-ARPES measurements were performed in our home laboratory at Tsinghua University using a Ti: sapphire regenerative amplifier producing femtosecond pulses at $\sim$800 nm ($\sim$1.55 eV) at 30 kHz repetition rate with an overall temporal resolution of $\sim$ 205 fs and energy resolution of $\sim$ 42 meV (see Fig.~S1 in the supplementary information). The infrared laser was frequency quadrupled to produce ultraviolet probe laser. The pump beam and probe beam were focused onto the sample with a FWHM (full width at half maximum) beam size of $\sim$220 $\mu$m and $\sim$120 $\mu$m, respectively. The tunable pump-probe delay was achieved by varying the pump optical path with a motorized delay stage of $\sim$6.7 fs precision. In order to minimize the space-charge effect \cite{Graf2010spacecharge}, the flux of the probe beam was lowered to make the shift and broadening of Fermi edge of the ARPES spectra negligible ($<$5 meV). Multi-photon photoemission from the pump beam was absent in our measured data for all chosen pump fluence. TR-ARPES measurements were performed on freshly cleaved MoTe$_2$ single crystals at 80 K with a base pressure better than 5$\times$10$^{-11}$ Torr.

The {\it ab-initio} calculations are carried out in the framework of the Perdew-Burke-Ernzerhof-type generalized gradient approximation of the density functional theory through employing the Vienna Ab initio simulation package (VASP) \cite{kresse1996} with the projected augmented wave (PAW) method. The {\bf k}-point mesh is taken as 12$\times$10$\times$6 for the bulk calculations. The spin-orbit coupling effect is self-consistently included.

\bibliographystyle{naturemaga}
\bibliography{TrMoTe2}

{\large{\bf Acknowledgments}}
This work is supported by the National Natural Science Foundation of China (Grant No.~11334006 and 11427903), Ministry of Science and Technology of China (Grant No. 2015CB921001 and 2016YFA0301004),  Science Challenge Project (No.~20164500122) and the Beijing Advanced Innovation Center for Future Chip (ICFC).



\end{document}